# POWER MARKET DYNAMICS: THE STATISTICAL MECHANICS OF TRANSACTION-BASED CONTROL

David P. Chassin*
Energy Science and Engineering Division
Pacific Northwest National Laboratory,
Richland, Washington 99352
United States of America

## ABSTRACT

Statistical mechanics provides a useful analog for understanding the behavior of complex adaptive systems, including power markets and the power systems they intend to govern. Transaction-based control is founded on the conjecture that the regulation of complex systems based on price-mediated strategies (e.g., auctions, markets) results in an optimal allocation of resources and an emergent optimal control. We outline a model based on strict analogies to thermodynamic quantities. The model accurately describes power market data collected from three North American independent system operators (ISO) in recent years. The ISO data is analyzed, comparing the behavioral similarities and differences that are observed.

## NOMENCLATURE

| | |
|---|---|
| $b$ | Power price basis [$/MWh] |
| $C$ | Cost [$/h] |
| $F$ | Transactive force [$·MW/h] |
| $g$ | System state degeneracy |
| $m$ | Machine fund reserves [$] |
| $M$ | Excess transaction moment [$·MW] |
| $n$ | Market elasticity coefficient |
| $N$ | Number of machines |
| $P$ | Price of a resource [$/MWh] |
| $p$ | Machine momentum [$·MW] |
| $Q$ | Power consumption [MW] |
| $U$ | Total value [$] |
| $x$ | Machine state |
| $Z$ | Grand sum |
| $e$ | Value [$] |
| $m$ | Unit intrinsic value [$] |
| $m_0$ | Transaction moment [$·MW] |
| $h$ | Supply elasticity |
| $s$ | System entropy |
| $t$ | Unit transactive value [$] |

* Corresponding author: Phone: +1 509 375 4369 Fax +1 509 375 3614 E-mail: david.chassin@pnl.gov

## INTRODUCTION

Physical models have been shown to have wide application to understanding the dynamics of complex adaptive systems [1]. Here we will show that it also may apply to the idea of controlling a complex engineered system using one or more market-like processes. The idea of using negotiation to resolve optimal control problems is not new, nor has it always been done with a literal market in the sense that price and quantity are the joint means of determining the allocation of a possibly scarce resource [2]. Nevertheless, the challenges faced by power market designers seem daunting. The complexities of these adaptive systems give rise to a multitude of emergent behaviors that frustrate engineers' attempts to design stable and robust classical control strategies [3,4]. Clearly, a well-defined set of quantities for the average properties of large dynamic systems is needed. Such properties would enable control engineers and system theorists to quantify both the time-independent and time-dependent dynamic behavior of large complex adaptive systems, thus permitting the design systems that are more likely to meet their requirement specifications and fail less frequently and less catastrophically.

## THE ABSTRACT MACHINE

We begin by defining an abstract machine and the system laws to which it would conform [5]. The purpose of this definition is twofold: 1) it provides a sufficiently general abstraction for all machines operating in market-based power systems; and 2) it provides a well-defined set of constraints on the behavior of machines, which enables closed-form derivation of many important average properties of systems composed of large numbers of these machines. (The statistical mechanical interpretation for the behavior of complex networks has been well enough established to be considered useful in this context [6].) This abstract machine is called an *Abstract Transactive Machine* and systems of these machines are referred to as *Transactive Control Systems.*

Figure 1 illustrates a machine that converts one or more resources, $Q_{in}$, (obtained at a cost $C_{out}$) into one or more resources, $Q_{out}$ (in consideration for which it receives $C_{in}$). For the sake of completeness, the machine may produce a byproduct $Q_{waste}$ for which no consideration or cost is incurred, and it may produce a profit $C_{profit}$ for which no resources are obtained.

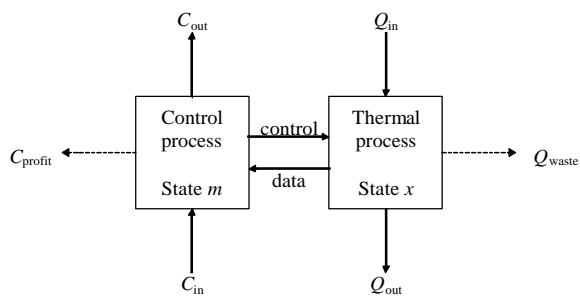

Figure 1: An abstract transactive machine

When two machines interact over the time $t$ of a transactive system, as shown in Figure 2, they exchange quantities $Qt$ of resources and $Ct$ of currency. In the simplest case, $Q$ and the price $P = C/Q$ are determined according to classical economics by evaluating the competitive price and quantity, which is the intersection of the supply and demand curves for the respective machines [7].

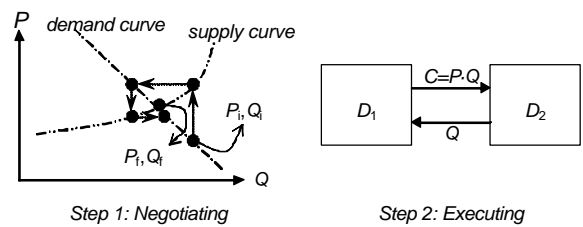

Figure 2: Two machines interacting

Transactions are the only interaction between machines that are permitted. Transactions occur in two steps. During the first step, an exchange of information between $D_1$ and $D_2$ takes place during which two important values are determined: the closing price $P$ at which the trade will occur and the contract quantity $Q$ of resource to be traded. The method of determining these values need not be known *a priori*, although one has been shown for illustrative purposes. An initial price and quantity $(P_i, Q_i)$ is offered or requested, for which a request or an offer is made, and negotiations continue until a final, or closing price and quantity $(P_f, Q_f)$ is determined and the first phase is concluded. During the second step, the agreed upon trade is actually performed and the values of $(P_f, Q_f)$ are in fact exchanged. This type of transaction is immediate. There are two other types of transactions that may also be negotiated: future and contingent. In a future transaction, the time at which the trade in step 2 will performed is negotiated. In a contingent transaction, the necessary conditions under which the trade will be executed are also negotiated. Consideration of these is beyond the scope of this paper.

However, in complex systems the supply and demand curves are not simply the summation of the individual machines' curves. They are also influenced by emergent topological phenomena, such as choices among multiple resources, the probabilities that machines occupy the requisites states to exploit those choices, and the ability of the underlying system to meet the delivery needs implicit in the trade. Because the particulars of these restrictions are unknown, we must constrain the behaviors of system such that they do not violate those restrictions. The restrictions must take the form of laws that express the known constraints preserved in the derivation of any quantity expressing an average property of the system.

We therefore postulate a series of dynamical laws that embody the recognized characteristics of machines and networks of machines (systems) and that will find expression in the derivation of any average property of the system. We do not suggest that these laws are currently enforced or even acknowledged in systems today, but they are intuitive and were they in force, the modeling of system and the detection of deviation from standard operation would be much simpler.

The proposed transactive system laws are:

*Law 1* Every machine has a state inertia, which is a joint effect of the internal thermodynamic process and the control process.

*Law 2* The effect of a machine on another machine's inertia is a joint function of its inertia and the rates of its thermodynamic process.

*Law 3* When two machines interact, they exchange equivalent values of resources and consideration.

*Law 4* The total value of a closed system remains constant.

Having postulated these laws, we may express them in a series of important dynamical relations among machines' properties. We define *momentum*, $p$ with respect to a given resource as the product of its cash reserves, $m$ (which is conceptually equivalent to particle mass in thermodynamics) and the rate of its thermodynamic processes, $Q$ (which is conceptually equivalent to velocity). Thus $p = m\,Q$ and a machine's *unit transaction moment* is $\boldsymbol{m}_0 \equiv dp = dm\,Q$. Because $C = dm/dt$, we therefore know that $\boldsymbol{m}_0 = P\,Q^2\,dt$. We define the *transactional force* of a machine as the instantaneous change in momentum, $F = dp/dt$ resulting from the action of another machine. We can therefore show that $F = m(Q^2/r + dQ/dt)$, where $r = m/P$, i.e., the relative value of a machine's cash reserves to the price of the resource. Finally, we shall define the total value $U$ of a system as the sum of its transactional value, its potential value, and its intrinsic value, or $U = U_T + U_P + U_I$. The transactional value $U_T$ represents the total value of the machines that is attributable to the interaction of the machines. The potential value $U_P$ is the unexploited part of the total value of the machines in the system. The intrinsic value $U_I$ is the total value based on the existence of the machine. Drawing on the analogy to statistical thermodynamics (which we will do throughout this paper), free value is that portion of the total value that is released as the system comes to equilibrium. Thus system operation involves an ever-changing redistribution of these values and system equilibrium is expected when the free value is minimum and entropy is maximum [8]. This is a completely analogous expression of Nash equilibrium in the sense that no further exchanges can be profitably performed between machines.

**SYSTEMS OF ABSTRACT MACHINES**

We can now define the formalism by which we represent machines as they come together to form systems and the markets that act on them. Markets provide a convenient representation of the collective effect of machines that are participating in transactions for a certain resource by aggregating supply and demand. By creating markets for multiple resources we create a $k$ dimensional space, one dimension for each resource providing one degree of freedom to the state of the machines.

Observations of electric power consumption suggest that machines exhibit approximate harmonic oscillatory behavior. Generation units often "swing" against each other in response to dynamic events on the transmission grid with characteristic frequencies [9]. Individual end-uses will consume resources at a rate that periodically increases and decreases in response to a constant supply [10]. (Indeed, it is noteworthy to observe that most of today's end-use technologies are completely unresponsive to the availability of supply, their control being based solely on the thermodynamic process governed in response to local prevailing conditions [11]. Thus, oscillatory behavior can exist in the absence of any influence by other machines. This observation is critical to the derivation of a truly statistical dynamic interpretation, which we know depends on the assumption that machines can explore all accessible states at any time.

**THE ELEMENTARY SYSTEM MODEL**

Having thus defined machines and systems, we may now formulate an elementary system model, from which, by strict analogy to the statistical treatment of thermodynamic systems, we shall derive important average properties of the system. The model system shown in Figure 3, is composed of abstract machines at fixed points on a line, each having transactive moment $\pm\boldsymbol{m}_0$ (+ for sellers, – for buyers). There are no interactions among the machines (i.e., no machine's state is as a consequence of another machine's state), and there is no external influence (i.e., no machine's state will change as a result of the action of an external machine). Each transactive moment may be oriented in one of two ways, sell (up) or buy (down), so that there are $2^8$ distinct arrangements of the eight transactive moments shown in the figure. If the arrangements occur in a random process, the probability of finding the arrangement shown is $1/2^8$.

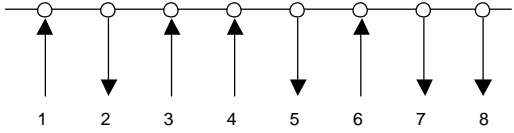

Figure 3: The elementary system model

Evaluating the N-factor product for the system, we follow the standard statistical mechanics analysis [12]. Then histogram of system state is the state degeneracy function $g$ for the system is

$$g(N,M) = g(N,0)e^{-2M^2/N} \qquad (1)$$

where $N$ is the number of machines in the system, $M$ is the excess transactive moment of the system, $1 \ll |M| \ll N$, and

$$g(N,0) \approx \frac{N!}{2\left(\frac{1}{2}N\right)!} \approx 2^N \left(\frac{2}{pN}\right)^{\frac{1}{2}} \qquad (2)$$

When two systems have excess transactive moments, $M_1$ and $M_2$ respectively, the total excess transactive moment $M$ of the combined system is $M_1 + M_2$. If they are in transactive contact with each other (meaning only value and not machines may move), then the combined value of these systems, $U(M)$ is $U_1(M_1) + U_2(M_2)$. Thus, we generalize the formulation of (1) over the accessible states of the combined system at the value $U$, and the degeneracy for the value of the combined system is

$$g(N,U) = \sum_{U_1=0}^{U} g_1(N_1,U_1)g_2(N_2,U-U_1) \qquad (3)$$

from which we obtain the formulation for the most probable configuration of the system. This is the configuration that satisfies the maximum value of the combined system:

$$\left(\frac{\partial \log g_1}{\partial U_1}\right)_{N_1} = \left(\frac{\partial \log g_2}{\partial U_2}\right)_{N_2} \qquad (4)$$

whence we derive the definition of the entropy of a system. Entropy $s$ is the logarithm of the degeneracy of its most probable configuration at the value $U$, or

$$s(N,U) \equiv \log g(N,U) \qquad (5)$$

where $s$ is maximal when the system is at equilibrium. It is using this definition of entropy that establishes the definition of transactive value $t$ in a system, as given by

$$\frac{1}{t} = \left(\frac{\partial s}{\partial U}\right)_N \qquad (6)$$

where $t$ is analogous to thermodynamic temperature and describes the value per machine of transaction-based activity in the system. When two systems are in transactive contact with each other, they are in transactive equilibrium if their respective transactive values, $t$ are equal.

We next derive the formulation for the intrinsic value, $m$ of a machine by considering the condition in which two systems are brought in diffusive contact, i.e., both value and machines are allowed to move between the systems while the combined system comes to equilibrium. Then the intrinsic value, $m$ is defined by the relation

$$-\frac{m}{t} \equiv \left(\frac{\partial s}{\partial N}\right)_U \qquad (7)$$

analogous to the definition of chemical potential in thermodynamic systems. Thus $U_I = N\,m$ and $U_T = N\,t$. When two systems in diffusive contact are in diffusive equilibrium, their respective transactive and intrinsic values are equal and the potential value difference between them is zero.

To determine the absolute probability that we will find the system in a given state we must first compute the normalizing factor needed to convert relative probabilities to absolute probabilities. The grand sum, $z$ over all states of the system for all numbers of machines is found to be given by

$$z(m,t) = \sum_{N=0}^{\infty} \sum_{l} e^{[Nm - e_l(N)]/t} \qquad (8)$$

where $e_l(N)$ is the value $e$ of the state $l$ when $N$ machines are present. We shall consider later a method of determining the value of the state of a machine. Thus the absolute probability of finding the system in a given state is thus given by

$$?(N,t) = \frac{e^{[Nm - e_l(N)]/t}}{z(m,t)}. \qquad (9)$$

We can also show that the average number of machines is given by

$$\langle N \rangle = l \frac{\partial}{\partial l} \log Z \qquad (10)$$

where $l \equiv e^{m/t}$.

Having derived these formulations for two average properties of transactive systems (i.e., transactive value and intrinsic value), it can be shown that a partition function $Z$ exists for any system, as given by

$$Z(N, t) = \sum_{l} e^{-e_l / t} \qquad (11)$$

From this summation over the Boltzmann factors, we can obtain the average value $U$ for a fixed number of machines:

$$U \equiv \langle e \rangle = \frac{\sum_l e_l e^{-e_l/t}}{Z(N,t)} = \frac{t^2}{Z}\frac{\partial Z}{\partial t} = t^2 \frac{\partial}{\partial t} \log Z \qquad (12)$$

## STATE DISTRIBUTION

Before going further, we must pause to consider the possibility that characterizing transactive system as fundamentally quantile is useful to developing a more robust model of system behavior. When considering the data collected on power end-uses, we observe that discrete cycling is common. Indeed, it is noteworthy to consider that many thermodynamic systems exhibit control hysteresis causing them to have a minimal cycling time. This is often necessary to protect equipment from the deleterious effects of fast cycling such as flooding of compressors. Furthermore, markets generally express marginal cost (hence the potential value of a transaction) as a step-wise function rather than a continuous one. We therefore propose that transactive systems composed of such machines are fundamentally quantile in nature. Thus, when a single two state machine (e.g., on or off) is in contact with a reservoir of transactive value $t$, then we have

$$\langle e \rangle = \frac{e e^{-e/t}}{Z} = e \frac{e^{-e/t}}{1 + e^{-e/t}} \qquad (13)$$

and

$$U = N \langle e \rangle = N e \frac{e^{-e/t}}{1 + e^{-e/t}} = \frac{Ne}{e^{e/t} + 1} \qquad (14)$$

In systems where the mode of one machine (i.e., the frequency and state path with which it conducts transactions) is not dependent upon that of another, so that state occupancy is non-exclusionary, strict analogy to the thermodynamic model leads us to conclude that the occupancy of a state is given by the Bose-Einstein distribution. Thus,

$$f(e) = \frac{1}{e^{(e-m)/t} - 1}. \qquad (15)$$

Indeed Bose-Einstein condensation phenomena have been previously observed in the behavior of web page links on the Internet [1].

In the case where state occupancy is exclusionary, it can be similarly shown that the grand sum is simply $Z = 1 + l e^{-e/t}$, and the average machine occupancy of the state $e$ is given by the Fermi-Dirac distribution [12], or

$$f(e) = \frac{1}{e^{(e-m)/t} + 1} \qquad (16)$$

To date, there are no examples in which the Fermi-Dirac distribution has been used to explain the behavior of complex systems. However, it is reasonable to conclude that expressions of market power, (e.g., withholding strategies) are discernable as state-exclusion behavior [7]. In addition, $q$-fermion distributions [13] permit finite state occupancies greater than 1 and may be considered when regarding the control strategies of power grid protection schemes, such as the staging of underfrequency load-shedding on distribution feeders designed to ensure that only the exact number of feeders occupy the requisite state to ensure system equilibrium [14].

## THE IDEAL SYSTEM MODEL

We now turn to the definition of an ideal transactive system in a classical regime, meaning one in which the condition of transactive value and concentration of machines in state space is such that the occupancy of any state is much less than 1. An ideal system is one in which the machines are free (no restrictions on momentum such as a price cap) and noninteracting (no changes in interdependence such as collusion) in the classical regime. We suggest that this ideal system model most closely approximates the behavior of systems in a topologically chaotic phase, when the average transaction duration between two machines is very close to the shortest possible transaction duration. The steady phase is one in which the average transaction duration approaches infinity, and the dynamic phase or so-called edge-of-chaos occurs when the average transaction duration is finite, but greater than the shortest one. Under these conditions, we can show that when $e - m \gg 1$, the two possible machine state distribution functions have nearly the same value, and therefore equations (15) and (16) are very nearly equal. Thus, we must have $e^{(e-m)/t} \gg 1$ for all $e$ and we find that

$$f(e) \cong 1 e^{-e/t} \quad (17)$$

which is the Maxwell-Boltzmann distribution function for the state of any transactive system in a classical regime.

Extensive study of the data collected on electric end-uses has revealed the discrete cycling that results from control hysteresis and other behavior of electric equipment [9]. But the only time we require definite knowledge of the state of a machine is at the beginning and end of transactions. Otherwise, it is sufficient that the machine does not violate any laws, allowing any number of transactions to occur between two boundary states (i.e., any whole number of definite intermediate states can occur between the end states is allowed). Except for those cases where so much of a resource is being consumed that a switch to another resource is likely, we can regard the potential value of a change in machine state as an infinitely deep well centered around the state having the minimum potential value (i.e., tunneling to another well is possible, but not likely). Hence we suggest that systems of such machines are quantized and the value of a machine in state $n$ is

$$e_n = (n+½)tw \quad (18)$$

where $t$ is the quantum transaction (in MWh$^2$) and $w$ is the frequency of transactions (per hour). That is, we assume that the potential value for which the machine can commit itself is, to lowest order, a one-dimensional quantum harmonic oscillator and the ground state, $e_0$ is the potential value of a machine that is plugged in but in a quiescent state. For example, such a quantized model of supply is readily recognized as a unit commitment curve, as illustrated in Figure 4.

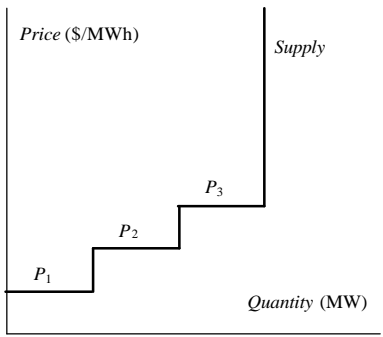

Figure 4: Unit commitment curve

Having thus found an expression for both the macroscopic and the microscopic values of systems and machines, we obtain what is for our present purposes the most important property of transactive systems, the distribution function for the rate $Q$. Using equations (17) and (18), this distribution can be shown to be given by

$$P(Q) = 4pN\left(\frac{m}{2pt}\right)^{\frac{3}{2}} Q^2 e^{-mQ^2/2t} . \quad (19)$$

It can be readily shown that the most probable rate is given by

$$Q_{mp} = \sqrt{\frac{2t}{m}} . \quad (20)$$

The mean rate is given by

$$\overline{Q} = \sqrt{\frac{8t}{p m}} \quad (21)$$

and the root mean-squared rate is given by

$$Q_{rms} = \sqrt{\frac{3t}{m}} . \quad (22)$$

It is by using these values that we can readily fit a model of an ideal transactive system to data collected from real systems. Thus we may solve equation (21) and (22) for the ratio of values of $t$ and $m$, from which we find

$$\frac{t}{m} = \frac{p}{8}\overline{Q}^2 = \frac{1}{3}Q_{rms}^2 = \frac{1}{2}Q_{mp}^2 \quad (23)$$

The market model may be found using a simple marginal cost curve that incorporates a notion of *scarcity rent* when supply approaches system capacity, such as the one given by

$$P = b\left(1 - \frac{Q}{\hat{Q}}\right)^{-1/n} \quad (24)$$

where $b$ is the price basis, $\hat{Q}$ is the system capacity, and $n$ is a market response coefficient. The supply elasticity $h = dQ/dP \cdot P/Q$ is given by

$$h = n\frac{1}{1 - P_{norm}^{-n}} \approx -n \quad (25)$$

for $P_{norm} = P/b \gg 1$ when scarcity rent is high. Starting with equation (24) we may perform a least squares fit for the parameters $A = -1/n$ and $B = \log b$ for the price data using the linear equation

$$\log P = A \log\left(1 - \frac{Q}{\hat{Q}}\right) + B. \qquad (26)$$

From this we may obtain estimates for the values of the price basis $b$ and the supply elasticity $h$.

## ANALYSIS OF ISO MARKET DATA

The average property of transaction-based control systems given in equation (19) is readily observable for many engineered systems. Indeed, we will compare the predictions of the rate distribution function to measurements taken of three major U.S. electric power independent system operators (ISOs), namely the Northeast (ISO-NE), Pennsylvania Jersey Maryland Interconnection, (PJM), and the California ISO (CAL-ISO). In the case of CAL-ISO we will study two different years to compare the results of the analysis.

### ISO New England

For ISO-NE we collected system load and energy price from the year 2001 shown in Figure 5. The price data plot indicates that the price was truncated at $250/MWh, but all the price data was used and the price cap based on observations of the data appears to be near $1000/MWh. We note that the price fluctuations do not always coincide with the seasonal fluctuations in load. Close inspection of the probability analysis of the load data (see Figure 6) indicates that the ISO-NE system has at least three distinct regimes.

This suggests that the data collected does not represent an entropic property of a single ideal system. This supposition is supported by Figure 7, where we find the price cap is reached for a wide range of loads between above 21 GW. In addition, the price cap is reached relatively frequently over a range of loads, suggesting that load and price are not well correlated, indicating a likely violation of the conservation laws required for an ideal system. Nevertheless, there is not evidence to suggest that there exist restrictions on action in markets such as those collusion might cause.

### Pennsylvania-Jersey-Maryland Interconnect

In the case of PJM we collected data for system load and locational-marginal price (LMP) for 1999, as shown in Figure 8. The load model shown reveals a better fit than that obtained for ISO-NE 2001. However the deviations in Figure 9 for probability of loads below 30 GW suggest that more than one regime may also be extant in the PJM system, although this phenomenon is not as pronounced as it is for ISO-NE.

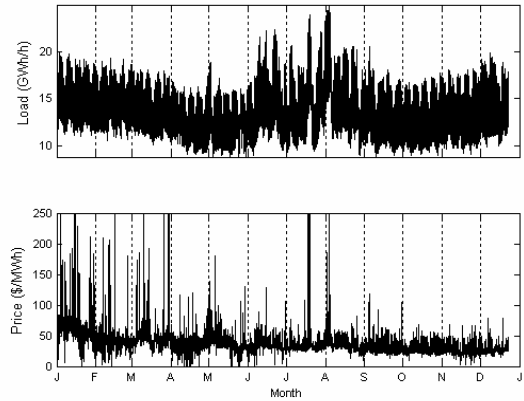

Figure 5: ISO-NE 2001 load and energy price

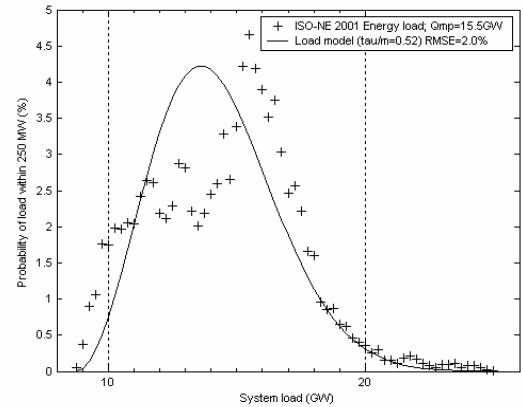

Figure 6: Load model of ISO-NE 2001

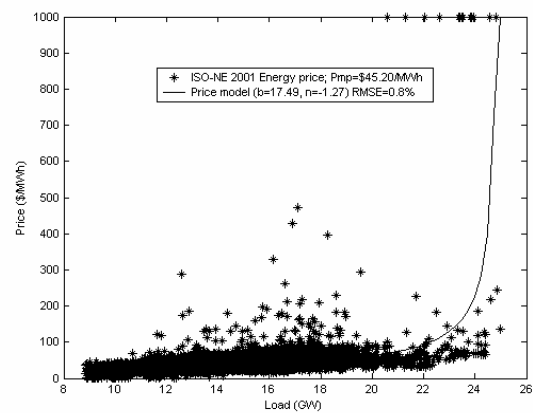

Figure 7: Market model of ISO-NE 2001

Indeed, inspection of the market model in Figure 10 reveals that either two distinct capacity regimes exist or one particular type of supply is not committed until the price goes above $900/MWh.

The former condition violates the prerequisite assumptions but the latter does not. For the ideal system model the dual regime is confirmed from the load history, and may be enough to explain the errors observed in the model fit at lower loads. Further analysis is required to understand this phenomenon.

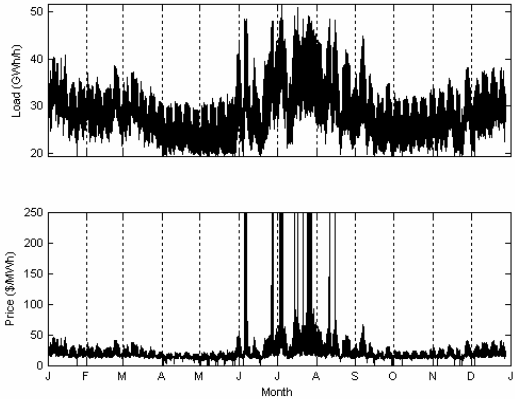

Figure 8: PJM 1999 system load and LMP

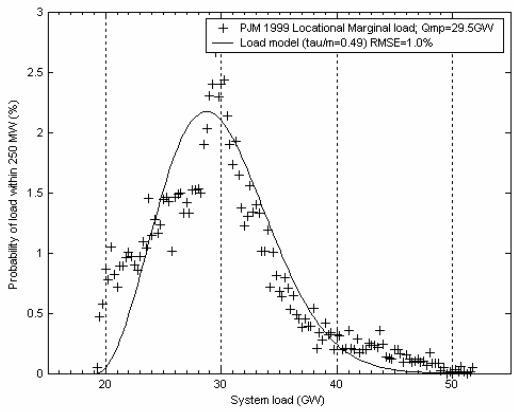

Figure 9: Load model of PJM 1999

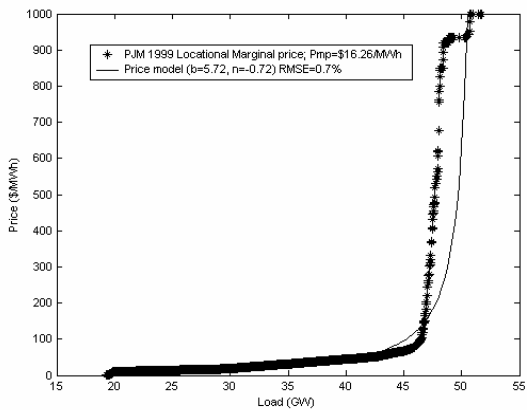

Figure 10: Market model of PJM 1999

**California ISO**

While we lacked price data for the California ISO for the years 2001 and 2002, we found it useful to study the load data shown in Figure 11 and Figure 12 to compare the models for the two years.

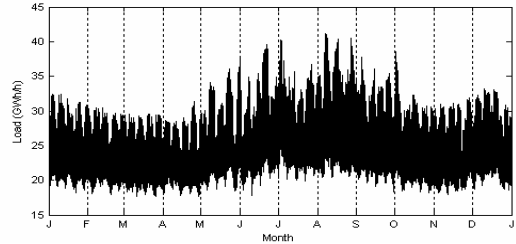

Figure 11: CAL-ISO 2001 system load

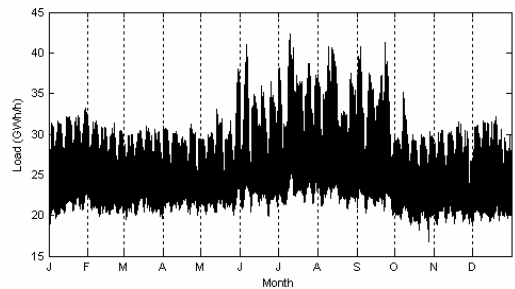

Figure 12: CAL-ISO 2002 system load

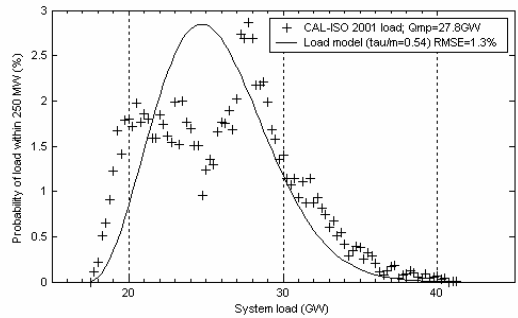

Figure 13: CAL-ISO 2001 load model

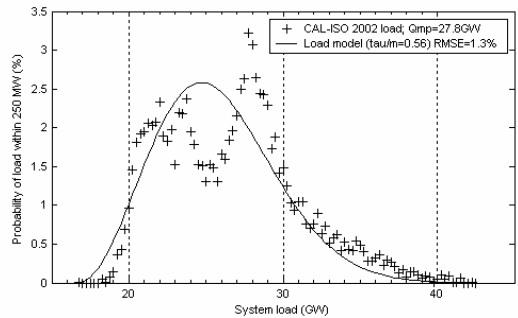

Figure 14: CAL-ISO 2002 load model

The overall appearance of the load data for the 2 years suggests two nearly identical systems. However, no hint of the strong multi-modal nature of the system seen in Figure 13 and Figure 14 is given by direct observation of the system load distribution. We speculate that these modes may be caused by diurnal or seasonal cycles, but it also possible that they are caused by a lack of economic interaction between transmission and distribution.

**Summary of Results**

Based on the derivation of the model given by equation (19) we interpret the parameters $t$ and $m$ to represent the value of interactions and the reserve value of machines, respectively. Therefore systems with higher $t$ have greater value by virtue of their activity than do systems with lower $t$. Similarly, systems with higher $m$ require more value to operate than systems with lower $m$.

|  | $t/m$ | $b$ | $h$ |
|---|---|---|---|
| **ISO-NE 2001 EP** | 0.52 | 17.49 | 1.27 |
| **PJM 1999 LMP** | 0.49 | 5.72 | 0.72 |
| **CAL-ISO 2001** | 0.54 | n/a | n/a |
| **CAL-ISO 2002** | 0.54 | n/a | n/a |

Table 1 shows a summary of the parameters determined for systems studied. From this we conclude that CAL-ISO is most active, and that PJM is operated with the least value. Furthermore, the fit $b$ and $n$ suggest that ISO-NE supply was significantly more expensive and less elastic than PJM, a conclusion supported by the fact the most probable prices were $45.20/MWh and $16.26/MWh for ISO-NE and PJM, respectively.

|  | $t/m$ | $b$ | $h$ |
|---|---|---|---|
| **ISO-NE 2001 EP** | 0.52 | 17.49 | 1.27 |
| **PJM 1999 LMP** | 0.49 | 5.72 | 0.72 |
| **CAL-ISO 2001** | 0.54 | n/a | n/a |
| **CAL-ISO 2002** | 0.54 | n/a | n/a |

Table 1: Model fit parameters

Tables 2, 3, and 4 provide a summary of the results for mean load, root mean square load, and most probable load. We see that the most significant errors occur for the most probable load. This seems largely because the systems considered are not strictly in a single classical regime, having multiple load frequency peaks over a wide range of loads and therefore more probably two or even three distinct systems or regimes, as noted above. This conclusion is supported by the fact the systems with fewer and less pronounced secondary peaks have much less error in predicting the most probable load.

**CONCLUSIONS**

The analysis of the data shows that the model proposed may have relevance to the systems being considered. The ability to derive a model from a combined first-principles consideration of both the economic and physical behaviors of machines opens many opportunities for enhancing system design and operation.

| System | Actual | Model | Error |
|---|---|---|---|
| **ISO-NE 2001 EP** | 14.5 | 14.3 | -0.2 |
| **PJM 1999 LMP** | 29.7 | 30.0 | 0.3 |
| **CAL-ISO 2001** | 26.0 | 25.7 | -0.3 |
| **CAL-ISO 2002** | 26.6 | 25.7 | -0.8 |

Table 2: Mean system load (GW)

| System | Actual | Model | Error |
|---|---|---|---|
| **ISO-NE 2001 EP** | 14.7 | 14.8 | 0.1 |
| **PJM 1999 LMP** | 30.3 | 30.9 | 0.6 |
| **CAL-ISO 2001** | 26.4 | 26.4 | 0.0 |
| **CAL-ISO 2002** | 26.9 | 26.5 | -0.4 |

Table 3: Root mean square system load (GW)

| System | Actual | Model | Error |
|---|---|---|---|
| **ISO-NE 2001 EP** | 15.5 | 13.7 | -1.8 |
| **PJM 1999 LMP** | 29.5 | 28.7 | -0.8 |
| **CAL-ISO 2001** | 27.8 | 24.8 | -3.0 |
| **CAL-ISO 2002** | 27.8 | 24.7 | -3.1 |

Table 4: Most probable system load (GW)

The multi-modal nature of the systems considered does not present an overwhelmingly difficult analytic challenge to using the model. Multi-modal analysis is extensively used in the characterization of many physical systems, and it seems quite reasonable to expect that those techniques will satisfactorily address the problems observed.

It is unclear at this time whether these analytic techniques will meet fully the requirements for detecting problems with market-based power systems, particularly with respect to early detection of market power abuse, overuse of price caps, or collusion. However, one can readily see the potential for such an application, particularly if a time-dependent model of system behavior can be developed based on this method.

It is interesting to speculate whether the exact state of machines need be known during the course of a transaction. Indeed, with respect to a given transaction, the machine is quite likely to conduct other transactions with other machines, perhaps even over the same resource. This uncertainty regarding the specific path and mode of a machine between two definite states that bound a transaction can be addressed by regarding the initial and final states of the machine as value barriers, requiring only that the machine not violate any laws, and allowing any number of other transactions to occur, creating one or more other determinate states between the initial and final state. Furthermore, it is likely that this quantized wave interpretation will eventually require us to recognize the existence of certain incompatible observables, such as might occur between transaction state and transaction rate.

This interpretation of machine state, combined with the potential value well model discussed previously suggests that the quantum harmonic oscillator model may be quite applicable to estimating the time-independent behavior of systems. We are particularly interested in the application of super-symmetry to the inverse scattering problem, according to which we may uncover families of partner potentials from the observations of market behavior [15]. These potentials may allow us to solve complex potentials by identifying simpler ones that have the same spectra. This may prove to be a very effective tool for system modeling and forecasting.

These results suggest that the abstract transactive machine and the transactive control system taken together are a useful model for understanding the behavior of certain complex adaptive systems. Clearly the assumptions made to enable this derivation make it likely that many systems cannot be accurately modeled at this time. However, it should be fairly easy to discern those systems to which this model can apply well. In the future, we expect work to eliminate some of these constraints to increase the number of complex adaptive systems that can be thus modeled and characterized. The day may indeed come when those models have the same effect on the design and operation of complex engineered systems that the introduction of the Carnot cycle had on the design and operation of thermodynamic systems.

My thanks to Jeffry V. Mallow of Loyola University Chicago's Department of Physics for his advice on this work. Pacific Northwest National Laboratory is operated by Battelle Memorial Institute for the U.S. Department of Energy under contract No. DE-AC06-76RL01830.